\documentclass{IEEEtran}
\usepackage{cite}
\usepackage{amsmath,amssymb,amsfonts}
\usepackage{algorithmic}
\usepackage{graphicx}
\usepackage{textcomp}
\usepackage{breqn}
\usepackage{amsmath}
\usepackage{multirow}
\def\BibTeX{{\rm B\kern-.05em{\sc i\kern-.025em b}\kern-.08em
    T\kern-.1667em\lower.7ex\hbox{E}\kern-.125emX}}
\usepackage{xparse}
\usepackage [latin1]{inputenc}
\NewDocumentCommand{\ceil}{s O{} m}{%
  \IfBooleanTF{#1} 
    {$\left\lceil#3\right\rceil$} 
    {#2\lceil#3#2\rceil} 
}

\begin{document}
\title{Multipath-Enhanced Measurement of Antenna Patterns: Theory}
\author{Daniel D. Stancil, \IEEEmembership{Life Fellow, IEEE}\thanks{The author is with the Department of Electrical and Computer Engineering, North Carolina State University, Raleigh, NC 27695 USA (e-mail: ddstanci@ncsu.edu). }}

\maketitle

\begin{abstract}
Traditional antenna pattern measurements involve minimizing the impact of multipath propagation in the measurement environment. In contrast, this work introduces a measurement approach that uses rather than mitigates multipath propagation. This is referred to as the Multipath-Enhanced Antenna Pattern (MEAP) Measurement technique. In this respect the approach has some kinship with Multiple-Input Multiple-Output (MIMO) systems. The advantage in the case of MIMO systems is increased capacity; in the MEAP approach the advantage is elimination of the need for creating an anechoic environment. The approach uses measurements with reference antennas to calibrate the multipath channel matrix, and vector spherical harmonics for efficient pattern representation. After presenting the mathematical details of the method, numerical calculations illustrating the approach are presented. Experimental results are described in a companion paper.
\end{abstract}

\begin{IEEEkeywords}
Antenna Pattern Measurements, Multipath Propagation, Vector Spherical Harmonics
\end{IEEEkeywords}

\section{Introduction}
\label{sec:introduction}
\IEEEPARstart{T}{he} traditional method of measuring antenna radiation patterns requires an anechoic environment so that there is a single known direction for test signals to be either sent or received \cite{hemming_electromagnetic_2002,Xu_anechoic_2019}. Consequently, multipath propagation was considered to be a source of errors, and techniques were developed to minimize its effects. Approaches explored include using the matrix pencil method to fit a sum of complex exponentials to the signal \cite{sarkar_using_1995,fourestie_anechoic_1999,loredo_echo_2004,leon_fernandez_radiation_2009}, time gating to keep only the direct path signal \cite{li_antenna_2001,loredo_echo_2004,loredo_measurement_2009,du_generation_2010,soltane_antenna_2020,olencki_low-cost_2023}, and various compensation and equalization techniques \cite{black_test_1995,leatherwood_plane_2001,leather_equalization_2003}.
However, anechoic chambers are expensive to construct, and a large number of pattern samples are often required to adequately characterize the three dimensional patterns. One way to reduce the cost of the facility is to use reverberation chambers. Reverberation chamber approaches include averaging out multipath using stirrers \cite{fiumara_free-space_2005,cozza_accurate_2010,ferrara_test_2012,fiumara_free-space_2016,puls_antenna_2018,reis_contactless_2022}, estimation of the Ricean K factor \cite{sorrentino_antenna_2013,lemoine_antenna_2013,besnier_radiation_2014}, and using the Doppler spectrum obtained by moving the antenna under test \cite{garcia-fernandez_antenna_2013,garcia-fernandez_antenna_2014}. 

In a somewhat analogous manner, multipath propagation in communication systems was traditionally  considered an impairment that must be mitigated by, for example, equalization or diversity combining. In contrast, following the work of Foschini and Gans \cite{foschini_limits_1998}, Telatar \cite{telatar_capacity_1999} and others, it was shown that if multiple transmit and receive antennas are used, multipath can be exploited to increase the information capacity of the system by creating multiple simultaneous channels. 

Here a similar point of view is applied to pattern measurements in non-anechoic environments. In particular, it is shown that the use of multiple measurement antennas can effectively realize multiple simultaneous channels, each of which conveys independent information about the antenna pattern.

To minimize the number of measurement antennas required, we need an efficient mathematical representation of the antenna pattern. A particularly compact way to represent 3D antenna patterns is with spherical harmonics \cite{hill_theory_1954,lambert_complete_1978,brock_using_2000,
allard_model-based_2003,ximenes_capacity_2010, marantis_comparison_2009,rahola_modelling_2009,mhanna_parametric_2012,schmitz_using_2012,mhedhbi_comparison_2013,fuchs_fast_2017,fuchs_compressive_2018,mezieres_application_2020,mezieres_fast_2021,mezieres_antenna_2021,huang_rapid_2022}. This work builds on these two ideas, using multipath to extract pattern information, and using spherical harmonics to minimize the number of sense antennas needed by efficiently representing the pattern information.

\section{System Configuration}
\begin{figure}[!t]
\centerline{\includegraphics[width=\columnwidth]{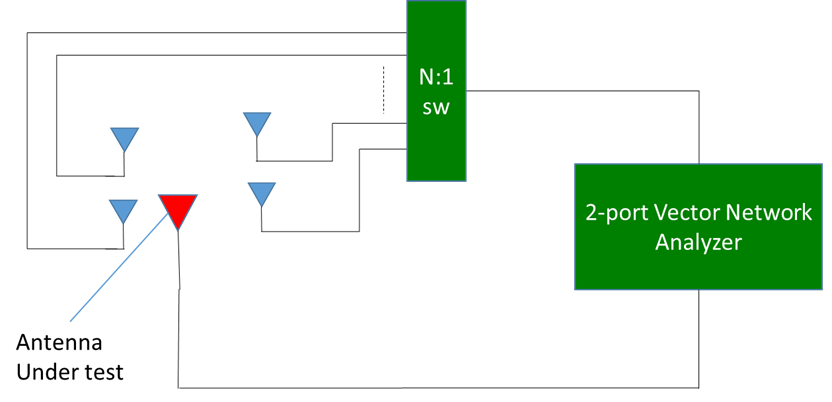}}
\caption{Diagram of the antenna measurement system. The antenna to be tested is placed in a multipath environment. Multiple probes are placed in the environment to sample the excited fields. An N:1 coaxial switch enables the 2-port vector network analyzer to measure the amplitude and phase of the signal coupled from each probe.}
\label{fig:system}
\end{figure}

A block diagram of the proposed system is shown in Figure \ref{fig:system}. The antenna under test is placed in a multipath environment. It is assumed that the scattering objects are sufficiently far from the antenna under test (AUT) that the presence of the sources of reflection do not significantly affect the behavior of the AUT. Example environments include an ordinary room, a large multimode waveguide, or reverberation chamber. The voltages induced at multiple probes placed in or around the surfaces of the chamber are then measured. The radiation properties are then obtained by analyzing the set of measured voltages. The analysis approach is described in the following sections.

The key idea is that the antenna pattern can be decomposed into a series of spherical harmonics, and each spherical harmonic pattern excites waves bouncing around in the environment in a unique way. By finding the mapping between spherical harmonics and the voltages measured at the probes, the complex amplitudes of each spherical harmonic in the series can be determined. The radiation properties of the antenna in free space can then be calculated based on the spherical harmonic decomposition. In the following sections these ideas are put on a more concrete mathematical basis.

\section{Spherical Harmonic Expansions}
Electromagnetic fields in a source-free region can be expanded in terms of vector spherical harmonics (VSH) \cite{hill_theory_1954,lambert_complete_1978,jackson_classical_1998,brock_using_2000} as well as scalar spherical harmonics (SSH) with vector coefficients \cite{rahola_modelling_2009,mhedhbi_comparison_2013}. Scalar spherical harmonics with vector coefficients have the advantage of avoiding singularities in the antenna field on the poles \cite{rahola_modelling_2009}, whereas Mhedhbi et al \cite{mhedhbi_comparison_2013} found that vector spherical harmonics gave a more efficient representation in terms of the number of harmonic coefficients needed. Since we are interested particularly in efficient pattern representations, we proceed with the vector spherical harmonic formulation.

For outgoing waves and assuming the time dependence $\exp{(j\omega t)}$ we have
\begin{multline}
    \mathbf{E} = \sum_{l,m} \left[a_{l,m}^M h_l^{(2)}(kr)\mathbf{X}_{l,m}\right. \\ - \left.\frac{j}{\omega \epsilon_0}a_{l,m}^E \nabla \times\left(h_{l,m}^{(2)}(kr)\mathbf{X}_{l,m}\right) \right],
\end{multline}
\begin{equation}
    \mathbf{H} = \frac{j}{\omega\mu_0}\nabla\times\mathbf{E}.
\end{equation}
Here the vector spherical harmonics are defined by
\begin{equation}
    \mathbf{X}_{l,m} = - \frac{j}{\sqrt{l(l+1)}}\mathbf{r} \times \nabla Y_{l,m} (\theta,\phi),
\end{equation}
and $h_l^{(2)}(x)$ is the spherical Hankel function
\begin{align}
    h_l^{(2)}(x) &= j_l (x)-jn_l(x)\\
    &=\sqrt{\frac{\pi}{2x}}\left[J_{l+1/2}(x)-jN_{l+1/2}(x)\right].
\end{align}
In the above, $Y_{l,m}$ is the spherical harmonic defined as
\begin{multline}
    Y_{l,m} (\theta,\phi)\equiv Y_{l,m}(\Omega) = \\ \sqrt{\frac{(2l+1)(l-m)!}{4\pi(l+m)!}}P_l^m(\cos\theta) e^{jm\phi}, \\ \begin{matrix} l=0,1,2,3...\\m=-l,-l+1...+l\end{matrix}
\end{multline}
$P_l^m(x)$ is the Associated Legendre function
\begin{equation}
    P_l^m(x) = (-1)^m(1-x^2)^{m/2}\frac{d^m}{dx^m}P_l(x),
\end{equation}
$P_l(x)$ is the Legendre function
\begin{equation}
    P_l(x) = \frac{1}{2^l l!}\frac{d^l}{dx^l}\left(x^2-1\right)^l,
\end{equation}
$j_l(x),\,n_l(x)$ are the spherical Bessel and Neumann functions, respectively, and $J_{l+1/2}(x),\,N_{l+1/2}(x)$ are the regular Bessel and Neumann functions of half-integer order.
Explicit expressions for the vector spherical harmonics are given in Appendix \ref{app:A}.

The vector spherical harmonics have the following properties \cite{lambert_complete_1978}:
\begin{align}
    \int \mathbf{X}_{l',m'}^* \cdot \mathbf{X}_{l,m}\text{d}\Omega &= \delta_{l,l'}\delta_{m,m'},\label{eq:ortho1}\\
    \int \mathbf{X}_{l',m'}^* \cdot \hat{\mathbf{r}}\times \mathbf{X}_{l,m}\text{d}\Omega &= 0,\label{eq:ortho2}\\
    \hat{\mathbf{r}}\cdot\mathbf{X}_{l,m} &= 0.\label{eq:ortho3}
\end{align}
Further, it follows from (\ref{eq:ortho1}) and (\ref{eq:ortho3}) that\footnote{Since $\hat{\mathbf{r}}$ is perpendicular to $\mathbf{X}_{l,m}$ from (\ref{eq:ortho3}), $\hat{\mathbf{r}}\times \mathbf{X}_{l,m}$ has a different direction but the same amplitude as $\mathbf{X}_{l,m}$. Consequently, $(\hat{\mathbf{r}}\times \mathbf{X}_{l',m'}^*)\cdot(\hat{\mathbf{r}}\times \mathbf{X}_{l,m})=\mathbf{X}_{l',m'}^* \cdot \mathbf{X}_{l,m}$, and (\ref{eq:ortho4}) follows.}
\begin{equation}
    \int (\hat{\mathbf{r}}\times \mathbf{X}_{l',m'}^*)\cdot(\hat{\mathbf{r}}\times \mathbf{X}_{l,m}) \text{d}\Omega = \delta_{l,l'}\delta_{m,m'}.\label{eq:ortho4}
\end{equation}

In the far-field, the large-argument expansion can be used for the spherical Hankel function $h_l^{(2)}(kr):$
\begin{equation}
    h_l^{(2)}(kr) \approx j^{l+1}\frac{e^{-jkr}}{kr},\quad kr>>1.
\end{equation}
It follows that the far-field approximations to the electromagnetic fields are
\begin{equation}
    \mathbf{E}=\frac{e^{-jkr}}{kr}\sum_{l,m}j^{l+1}\left[a_{l,m}^M \mathbf{X}_{l,m}-\eta_0\, a_{l,m}^E\hat{\mathbf{r}}\times\mathbf{X}_{l,m}\right],
    \label{eq:E_expand}
\end{equation}
\begin{equation}
    \mathbf{H} =\frac{1}{\eta_0}\hat{\mathbf{r}}\times\mathbf{E}.
\end{equation}
Here $\eta_0 = \sqrt{\mu_0\epsilon_0}$ is the impedance of free space. 

Alternatively, the coefficients can be extracted from a measured field $\mathbf{E}$ using
\begin{equation}
    a_{l,m}^M = (-j)^{l+1}kr\,e^{jkr}\int \mathbf{E}\cdot\mathbf{X}_{l,m}^*\text{d}\Omega,
\end{equation}
\begin{equation}
    a_{l,m}^E = -\frac{(-j)^{l+1}}{\eta_0}kr\,e^{jkr}\int\mathbf{E}\cdot\left(\hat{\mathbf{r}}\times\mathbf{X}_{l,m}^*\right)\text{d}\Omega.
    \label{eq:AE_coeff}
\end{equation}
The coefficients for both electric and magnetic dipole components have the property
\begin{equation}
    a_{l,-m}=(-1)^m a_{l,m}^*.
    \label{eq:coef_sym}
\end{equation}
The complex Poynting vector is given by
\begin{equation}
    \mathbf{S} = \frac{1}{2}\mathbf{E}\times\mathbf{H}^* = \frac{1}{2\eta_0}\mathbf{E}\times(\hat{\mathbf{r}}
    \times\mathbf{E}^*) = \frac{1}{2\eta_0}|\mathbf{E}|^2\hat{\mathbf{r}}.
    \label{eq:poynting}
\end{equation}
The last step follows since $\mathbf{E}\cdot\hat{\mathbf{r}}=0$ in the far-field. The magnitude squared of the electric field can be written
\begin{align}
    \begin{aligned}
        |\mathbf{E}|^2 &= \frac{1}{(kr)^2}\sum_{l',m'}j^{l'+1}\sum_{l,m}(-j)^{l+1}
        \left[a_{l',m'}^M\mathbf{X}_{l',m'}\right.\\
        &\quad\left.-\eta_0 \,a_{l',m'}^E\hat{\mathbf{r}}\times\mathbf{X}_{l',m'}\right]^*\cdot\left[a_{l,m}^M\mathbf{X}_{l,m}-\eta_0 \,a_{l,m}^E\hat{\mathbf{r}}\times\mathbf{X}_{l,m}\right]\\
        &=\frac{1}{(kr)^2}\sum_{l',m'}j^{l'+1}\sum_{l,m}(-j)^{l+1} \left[(a_{l',m'}^M)^*a_{l,m}^M\mathbf{X}_{l',m'}^*\cdot\mathbf{X}_{l,m}\right.\\
        &\quad +\eta_0^2(a_{l',m'}^E)^*a_{l,m}^E(\hat{\mathbf{r}}\times\mathbf{X}_{l',m'}^*)\cdot(\hat{\mathbf{r}}\times\mathbf{X}_{l,m})\\
        &\quad\left.-\eta_0\left((a_{l',m'}^M)^*a_{l,m}^E\mathbf{X}_{l',m'}^*\cdot\hat{\mathbf{r}}\times\mathbf{X}_{l,m} + c.c\right)\right]
        \label{eq:E2}
    \end{aligned}
\end{align}
To find the total power radiated, we substitute this into (\ref{eq:poynting}) and integrate over the surface of a sphere at the far-field radius $r$.
The last two terms in (\ref{eq:E2}) vanish owing to the orthogonality relation (\ref{eq:ortho2}), and the first two terms simplify using the orthonormal relations (\ref{eq:ortho1}) and (\ref{eq:ortho4}). The result is
\begin{equation}
    P = \frac{1}{2\eta_0 k^2}\sum_{l,m}\left[|a_{l,m}^M|^2 + \eta_0^2|a_{l,m}^E|^2\right].
    \label{eq:radpwr}
\end{equation}

The radiation resistance of the antenna is then given by
\begin{equation}
    R_{\text{r}} = \frac{2P}{|I|^2},
    \label{eq:radres}
\end{equation}
where $I$ is the current input to the antenna terminals.

For antennas with high efficiency, the radiation resistance and the antenna impedance (assuming no reactive part) will be very nearly equal. In this case the antenna impedance is more readily measured with a vector network analyzer connected to the antenna terminals (to measure $S_{11}$). Eq. (\ref{eq:radres}) can then be compared with the impedance obtained from the network analyzer as a check to ensure that the spherical harmonic expansion is correct. For low efficiency antennas, the efficiency of the antenna can be estimated by comparing the radiation resistance (\ref{eq:radres}) to the antenna terminal impedance.

Finally, the directivity can be obtained from
\begin{equation}
    D = \frac{4\pi r^2}{P}\frac{\max{|\mathbf{E}}|^2}{2\eta_0}.
\end{equation}

For calculations, it is convenient to introduce a ``modified field" and ``modified electric multipole coefficient" defined as
\begin{align}
    \tilde{\mathbf{E}} & = kr\,e^{jkr}\,\mathbf{E},\\
    \tilde{a}_{l,m}^E &= -\eta_0\,a_{l,m}^E.
\end{align}
In terms of these quantities we have
\begin{equation}
    \tilde{\mathbf{E}} = \sum_{l,m}j^{l+1}\left[a_{l,m}^M\mathbf{X}_{l,m}+\tilde{a}_{l,m}^E \hat{\mathbf{r}}\times\mathbf{X}_{l,m}\right],
\end{equation}
\begin{equation}
    a_{l,m}^M = (-j)^{l+1}\int{\tilde{\mathbf{E}}\cdot\mathbf{X}_{l,m}^*\text{d}\Omega},
    \label{eq:Mcoef}
\end{equation}
\begin{equation}
    \tilde{a}_{l,m}^E = (-j)^{l+1}\int{\tilde{\mathbf{E}}\cdot (\hat{\mathbf{r}}\times\mathbf{X}_{l,m}^*)\,\text{d}\Omega},
    \label{eq:Ecoef}
\end{equation}
\begin{equation}
    P = \frac{1}{2\eta_0 k^2}\sum_{l,m}\left[|a_{l,m}^M|^2 + |\tilde{a}_{l,m}^E|^2\right],
\end{equation}
\begin{equation}
    D = \frac{4\pi\max{|\tilde{\mathbf{E}}|^2}}{2\eta_0k^2P}.
\end{equation}

To summarize, if we know the coefficients of the vector spherical harmonics of the radiated field $a_{l,m}^E,\,a_{l,m}^M$ for a given $I$, then we can calculate the radiation pattern, directivity, and radiation resistance of the antenna.

\section{Antenna in a Multipath Environment}
Suppose that an antenna to be tested is placed in a multipath environment. If there are $N_s$ sensing antennas and $N_p$ significant paths, then the voltages induced at the $m$th antenna can be written
\begin{equation}
\begin{aligned}
    v_m &= \sum_{n=1}^{N_p}Q_{mn}c_n,\\
   \mathbf{v} &= \mathbf{Q}\cdot
    \mathbf{c}.
\end{aligned}
\end{equation}
Here $\mathbf{Q}$ is a $N_s \times N_p$ coupling matrix between the sensing antennas and the multiple paths, and $\mathbf{c}$ is a $N_p\times\,1$ matrix (column vector) containing the amplitudes of the paths.

Now suppose that the antenna under test is band-limited in spherical harmonics; i.e., there is a finite number $N_\Lambda$ of vector spherical harmonics with non-negligible amplitudes. Here the band-limited assumption is imposed by assuming that all coefficients have negligible amplitudes for $l>\Lambda$. For each value of $l$ there are $2l+1$ spherical harmonics. This observation applies to both $a_{l,m}^E,a_{l,m}^M$, so the limit on the total number of harmonics for the band-limited system is therefore
\begin{equation}
    N_\Lambda = 2\sum_{l=1}^\Lambda (2l+1) = 2\Lambda(\Lambda+2).\label{eq:Nmodes}
\end{equation}
Note that we have started the sum with $l=1$ since for antenna problems, the spherically symmetric function for $l=0$ will not occur. 

Each one of these spherical harmonic components will excite a unique distribution of paths, so we can write
\begin{equation}
    \mathbf{c} = \mathbf{D}\cdot\mathbf{a},
\end{equation}
where
\begin{equation}
    \mathbf{a} = \begin{bmatrix}\mathbf{a}^E\\\, \\\mathbf{a}^M\end{bmatrix},\quad \mathbf{a}^E = \begin{bmatrix}a_{1,-1}^E\\ \,\\a_{1,0}^E\\\vdots\\\, \\a_{\Lambda,\Lambda}^E\end{bmatrix},\quad \mathbf{a}^M = \begin{bmatrix}a_{1,-1}^M\\ \,\\a_{1,0}^M\\\vdots\\\, \\a_{\Lambda,\Lambda}^M\end{bmatrix}.
\end{equation}
Here $\mathbf{D}$ is an $N_p\times N_\Lambda$ matrix describing the coupling between vector spherical harmonics and propagation paths, and $\mathbf{a}$ is an $N_\Lambda\times 1$ matrix (column vector) of vector spherical harmonic amplitudes associated with the antenna under test. The connection between the vector spherical harmonic amplitudes and the measured probe voltages can therefore be written
\begin{equation}
    \mathbf{v} =\mathbf{Q}\cdot\mathbf{D}\cdot\mathbf{a}.
\end{equation}
If $\mathbf{Q}\cdot\mathbf{D}$ is invertible, then the spherical harmonic amplitudes can be recovered from the voltage measurements using
\begin{equation}
    \mathbf{a}=\left[\mathbf{Q}\cdot \mathbf{D}\right]^{-1}\cdot \mathbf{v}.
    \label{eq:a_calc}
\end{equation}
Note that to preserve information, both the number of paths and the number of sensing antenna measurements must be at least as large as the number of non-negligible vector spherical harmonics; i.e., $N_p\ge N_\Lambda,\,N_s\ge N_\Lambda$.

\section{Calibration of the Environment}
The matrices $\mathbf{Q}$, $\mathbf{D}$ could be computed from first principles using the geometry of the environment, the equivalent antenna currents, and the probe geometry, but experimentally it may be preferable to calibrate the system from a series of reference antennas. (The first-principles calculation would likely need to be numerical in general, depending on the complexity of the environment.) A reference antenna in this context would be one for which the complete pattern has been measured or computed and decomposed into its vector spherical harmonic components. 

Given $N_R$ reference antennas, we can construct the $N_\Lambda\times N_R$ matrix whose columns are the spherical harmonic coefficient vectors for each reference antenna:
\begin{equation}
    \mathbf{A}_R = \begin{bmatrix} \mathbf{a}_1 & \mathbf{a}_2 & \cdots & \mathbf{a}_{N_R}\end{bmatrix}.
\end{equation}
If we measure the sensing antenna voltages for each of these reference antennas we can also construct the $N_s\times N_R$ voltage matrix $\mathbf{V}$ such that
\begin{equation}
    \mathbf{V}_R = \begin{bmatrix} \mathbf{v}_1 & \mathbf{v}_2 & \cdots & \mathbf{v}_{N_R}\end{bmatrix}.
\end{equation}
It follows that
\begin{equation}
    \mathbf{V}_R = \mathbf{Q} \cdot\mathbf{D} \cdot\mathbf{A}_R. 
\end{equation}
If the matrix $\mathbf{A}_R$ is invertible, then the matrix $\mathbf{Q} \cdot \mathbf{D}$ can be obtained from
\begin{equation}
    \mathbf{Q} \cdot \mathbf{D} = \mathbf{V}_R \cdot \mathbf{A}_R^{-1},
    \label{eq:QD_calc}
\end{equation}
and
\begin{equation}
    \left[\mathbf{Q} \cdot \mathbf{D}\right]^{-1} = \mathbf{A}_R \cdot \mathbf{V}_R^{-1}.
\end{equation}

It is also possible to separate the analysis into two steps as follows. Let $v_j^{TA}$ be the measured voltage at the $j$th sensing antenna owing to the antenna under test. An estimate $v_j^{e}$ of $v_j^{TA}$ can be constructed with a weighted sum of the voltages from the reference antennas:
\begin{equation}
    \mathbf{v}^e = \mathbf{V}_R\cdot \mathbf{w}.
\end{equation}
In this case the weights can be obtained directly from
\begin{equation}
    \mathbf{w} = \mathbf{V}_R^{-1}\cdot \mathbf{v}^e.
\end{equation}
Owing to the linearity of the system, an estimate of the vector containing the spherical harmonic amplitudes of the test antenna would then be
\begin{equation}
    \mathbf{a}^e = \mathbf{A}_R\cdot \mathbf{w}.
\end{equation}
If we require that all antennas are mounted so that their phase centers are coincident, then the coefficients $w_i$ should be real. 

\section{Least-Square-Error Method}
The calculations in the previous section require key matrices to be invertible, namely square and non-singular. While squareness can be achieved by choosing $N_s = N_\Lambda$, singular or ill-conditioned matrices will occur if the amplitudes of some spherical harmonics are zero or close to zero. Since we assume we do not know the amplitudes a priori, we need an approach that avoids the need to invert ill-conditioned matrices. The least-square-error (LSE) method should be robust independent of the spherical harmonic content, so we proceed with this approach \cite{mhedhbi_comparison_2013}.

The formulation of the LSE method is as follows. 
We would like to find the weights $w_i$ that minimize the error function
\begin{equation}
    \varepsilon = \left(\mathbf{v}^{TA}-\mathbf{v}^e\right)^\dagger \cdot \left(\mathbf{v}^{TA}-\mathbf{v}^e\right),\label{eq:costfun}
\end{equation}
where $\mathbf{v}^{TA}$ is the vector of measured voltages for the test antenna.
 
Setting the derivatives with respect to $w_i$ to zero gives a set of simultaneous equations that can be written in matrix form and solved for the coefficients $w_i$:
\begin{equation}
  \mathbf{w} = \left[\text{Re}\{{\mathbf{V}_R}^\dagger\cdot \mathbf{V}_R\}\right]^{-1}\cdot \text{Re}\{{\mathbf{V}}_R^\dagger\cdot {\mathbf{{v}}}^{TA}\},
\end{equation}
where $\mathbf{w}$ is the vector of coefficients $w_i$.

It may be possible to improve the accuracy by imposing a normalization condition as a constraint on the minimization of (\ref{eq:costfun}). Such a normalization is equivalent to holding the total power constant. A straightforward choice is simply
\begin{equation}    
    \mathbf{w}^{\text{T}}\cdot\mathbf{w} = 1.
\end{equation}
If the efficiency of the test antenna is known, another choice of normalization is the radiation resistance constraint (\ref{eq:radres})
\begin{equation}
    \frac{2P}{|I|^2} = R_{\text{Meas}} - R_{\text{loss}},
\end{equation}
where $R_{\text{Meas}}$ is the actual measured terminal resistance, and $R_{\text{loss}}$ represents the loss of the antenna in the form of a resistance in series with the radiation resistance. (We assume reactive components have been tuned out.)


\section{Choice of Model Parameters}
\label{sec:parameters}

\subsection{Estimating the Number of Modes}
For a given spherical harmonic with index $l$, the mode decays rapidly with increasing $r$ if $kr<l$, and decays as $1/r$ for $kr>l$. \cite{cappellin_properties_2008}. Consequently, the mode is evanescent for $kr<l$, and propagating for $kr>l$.
It follows that if the smallest sphere that encloses the antenna has the radius $R$, then the contributions from spherical harmonics with $l>kR$ rapidly decrease with increasing $l$.


Thus an estimate for the limit on $\Lambda$ is given by
\begin{equation}
    \Lambda \leq \ceil[\Big]{kR} = \ceil[\Big]{\frac{2\pi R}{\lambda}},
    \label{eq:Lambda1}
\end{equation}
where $\lceil\, \rceil$ represents the ceiling function.

Jensen and Frandsen \cite{jensen_number_2004} have studied the amount of relative power excluded by truncating the spherical harmonic series, and found that for an infinitesimal dipole located at $R$ such that $kR=30$, truncation at $\Lambda = kR$ results in excluding relative power of about -15 dB. Based on additional calculations for larger values of $kR$, they gave a more conservative estimate based on the desired level of accuracy:
\begin{equation}
    \Lambda = kR + 0.045\sqrt[3]{kR}(P_R - P_{tr}),
\end{equation}
where $P_R$ is the power in dB of the source at $r=R$ relative to the total power, and $P_{tr}$ is the power excluded by the truncation, relative to $P_R$ (dB). Taking $P_R=0$ dB gives
\begin{equation}
    \Lambda = kR - 0.045 \sqrt[3]{kR}P_{tr}.
    \label{eq:Lambda3}
\end{equation}
This expression is valid for $kR\gg 1$ and $P_{tr}\leq -40$ dB \cite{jensen_number_2004}. For example, for $P_{tr}=-40$ dB, we have
\begin{equation}
    \Lambda = kR +1.8\, \sqrt[3]{kR}.
\end{equation}

Once an estimate for $\Lambda$ is obtained, the resulting number of spherical harmonics is given by (\ref{eq:Nmodes}). This assumes that the antenna could have both electric and magnetic multipole sources. If it is known, for example, that the antenna only has electric or magnetic multipole sources, (e.g., a loop antenna or an electric dipole), then the number of coefficients is reduced by half.

For small antennas where $kR\ll 30$, it may be necessary to begin with one of these estimates, then change the number of VSH amplitudes included until the results satisfy the required accuracy.

In addition to the reduction of complexity realized by estimating the band-limited properties of the pattern, it may often be the case that many spherical harmonic coefficients have negligible amplitudes even within the band-limited range. Thus, even within the band-limited constraint, the coefficient vector may be sparse. In this case, it is likely that compressed sensing techniques \cite{candes_introduction_2008} can be used to minimize the number of measurements needed to obtain accurate reconstruction of the function.

\subsection{Choice of Sensor Antenna Locations }
If the matrix inversion method is used, then a key requirement is for the matrix $\mathbf{Q}\cdot \mathbf{D}$ to be inverted, i.e., to be well conditioned. Consequently, the locations of the sensing antennas as well as the set of reference antennas should be chosen such that the condition number of $\mathbf{Q}\cdot \mathbf{D}$ is minimized.

More generally, each distinct multipath component carries independent information about the pattern (and consequently the VSH composition) of the antenna. Thus we can view the preferred sensing antenna locations as those that maximize the information capacity of the channel matrix. If we view the vector of VSH amplitudes $\mathbf{a}$ as the signal transmitted by the antenna, and the set of voltages $\mathbf{v}$ received by the sensing antennas as the received signal, then we can define an effective channel matrix as $\mathbf{T}\equiv \mathbf{Q}\cdot \mathbf{D}$. The $\epsilon$-capacity is the maximum information that can be transmitted through the channel matrix $\mathbf{T}$ in the presence of an uncertainty level of $\epsilon$. The $\epsilon$-entropy gives an upper bound to the $\epsilon$-capacity, and is given by \cite{behjoo_optimal_2022}:
\begin{equation}
    H_\epsilon (\mathbf{T}) = \frac{1}{2}\log_2\left(\frac{\det{(\mathbf{T}\cdot\mathbf{T}^\dagger)}}{\epsilon^2}\right) = \sum_{k=1}^{N_s}\log_2\left(\frac{\sigma_k}{\epsilon}\right),
\end{equation}
where $\sigma_k$ is the $k$th singular value of $\mathbf{T}$. Regardless of the value of $\epsilon$, the upper bound on the capacity will be proportional to $H_1$. It is therefore desirable to maximize $H_1$, or equivalently to minimize:
\begin{equation}
    f(\overline{T}) = -\det{(\mathbf{T}\cdot\mathbf{T}^\dagger)}.
\end{equation}
Beyond maximizing $H_1$ or minimizing $f$, it is important to know whether or not the optimum value represents a matrix with sufficient degrees of freedom to approximate the VSH content with the required accuracy. It is likely that the Kolmogorov $n$-width can be used to obtain a quantitative estimate of the degrees of freedom in this context \cite{migliore_electromagnetics_2008}, but this requires additional exploration.

Note that if $\mathbf{A}_R$ is well conditioned, we could also consider the different reference antennas to be the inputs to a multiple-input multiple-output (MIMO) system (imagining that the reference antennas could be simultaneously energized at the same phase center location). In this case, $\mathbf{V}_R$ could be considered to play the role of the channel matrix. However, we are ultimately interested in how the VSH coefficients of the AUT map onto the measured sensor voltages, and the matrix $\mathbf{Q}\cdot\mathbf{D}$ takes into account both the diversity of the environment and the VSH diversity of the set of reference antennas.

\section{Example: Half-wave Dipole with Arbitrary Orientation in a Random Multipath Environment}
\subsection{Half-Wave Dipole with Arbitrary Orientation}
Consider the case of antennas that are center-fed dipoles of various lengths and orientations. The geometry is shown in Figure \ref{fig:dipole}. Explicit expressions for the electric field are given in Appendix \ref{app:B}.
As a simple example, suppose that we limit our test antennas to half-wavelength dipoles (i.e., $L=\lambda/2$), but at some arbitrary orientation. Using the simplest estimate of  $\Lambda$ from Eq. (\ref{eq:Lambda1}) gives $\Lambda=2,N_\Lambda=16$, since the smallest radius enclosing the antenna would be $R=L/2=\lambda/4$. In contrast, the more conservative estimate (\ref{eq:Lambda3}) gives $\Lambda=4,N_\Lambda=48$. However, since we know that this antenna only has electric multipole equivalent sources, we could reduce our estimate to $N_\Lambda\approx 8$ or $N_\Lambda\approx 24$. In the case of the half-wave electric dipole, the pattern symmetry allows us to further reduce the required number of coefficients. In particular, the coefficients vanish for even values of $\Lambda$, so it is sufficient for us to consider the case $\Lambda=3$.

\begin{figure}[!t]
\centerline{\includegraphics[width=0.7\columnwidth]{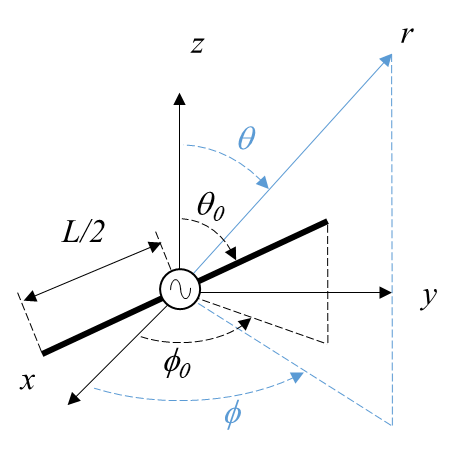}}
\caption{Geometry for a center-fed, symmetric electric dipole antenna. The dipole is centered at the origin with its axis along the direction $(\theta_0,\phi_0)$, and the observation is made at a distance $r$ in the direction $(\theta,\phi)$.}
\label{fig:dipole}
\end{figure}

\begin{figure}[!t]
\includegraphics[width=3.5in]{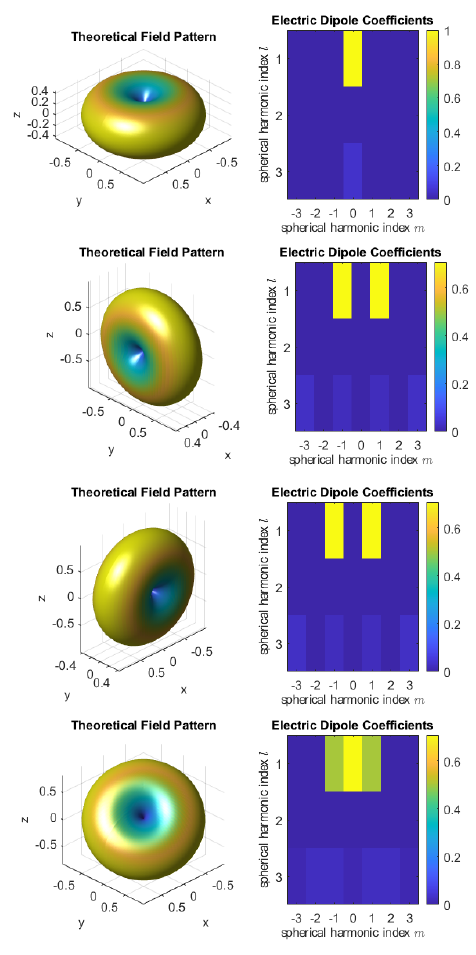}
\caption{Examples of spherical harmonic decomposition of half-wave dipoles with various orientations. The coefficients for $\pm m$ have the same magnitudes but different phases (phases not shown). Orientations from top to bottom: ($\theta,\phi$) = (0,0), ($\pi/2,0$), ($\pi/2,\pi/2$), ($\pi/4,\pi/4$). Coefficient magnitudes are normalized to the top case (0,0).}
\label{fig:dipole_examples}
\end{figure}

Figure \ref{fig:dipole_examples} shows the normalized patterns and spherical harmonic spectra for $\Lambda=3$ for several orientations. Referring to the spectra, we see that the significant terms $(l,m)=(1,0),\,(1,\pm 1)$ are indeed covered if we were to take $\Lambda=2$ as given by Eq. (\ref{eq:Lambda1}). In this case only 3 spherical harmonic amplitudes are needed, and the top three orthogonal orientations in Fig.  \ref{fig:dipole_examples} can be used as calibration antennas. 

However to ensure a more accurate representation, we do see some contributions from $(l,m)=(3,0),\,(3,\pm 1),\,(3,\pm 2),\,(3,\pm 3)$ as well. Since the symmetry of the dipole patterns results in no components for $l=2$, the total number of spherical harmonic amplitudes we need for $\Lambda=3$ is 10.\footnote{In view of the relation between coefficients with values of $\pm m$ given by (\ref{eq:coef_sym}), it may appear that the actual number of coefficients is smaller. However, counting the real and imaginary parts of a coefficient as independent unknowns, there are equivalently 10 real unknowns to be determined.}

\begin{figure*}[!t]
\includegraphics[width=\textwidth]{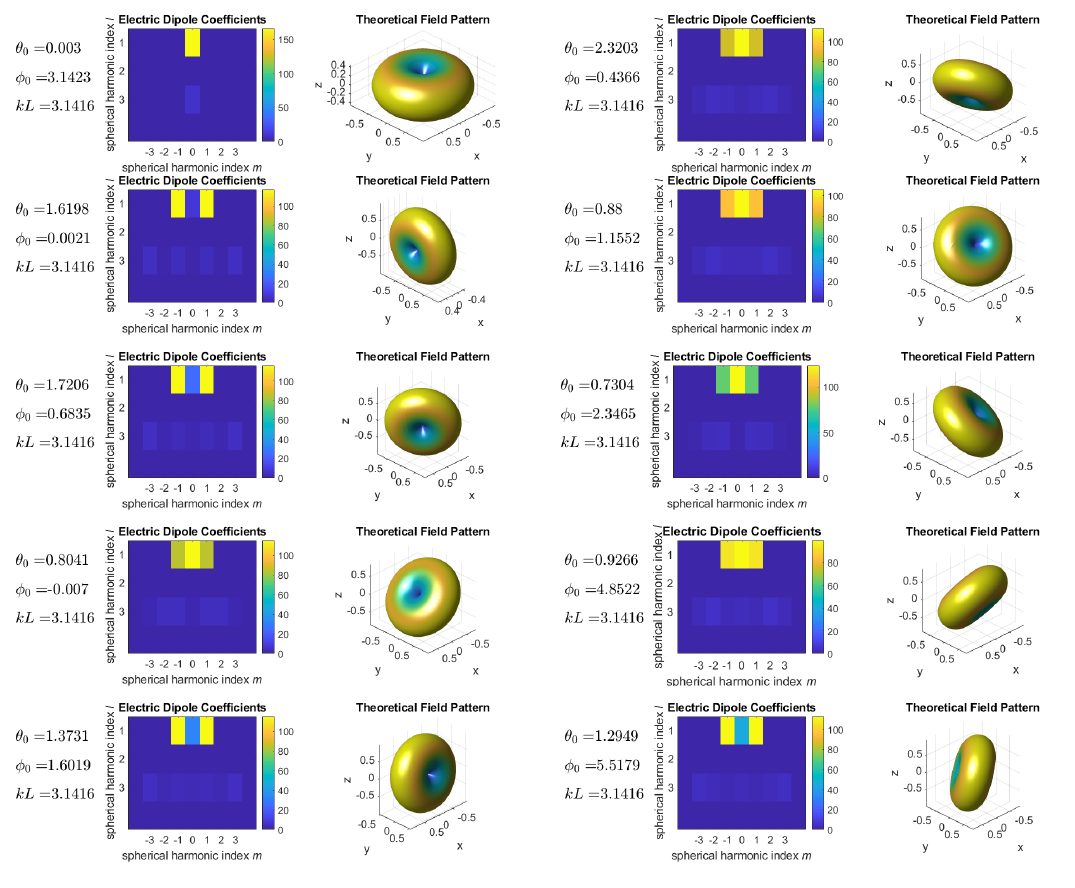}
\caption{Properties of a set of electric dipole calibration antennas.}
\label{fig:ref_dipoles}
\end{figure*}

To obtain the amplitudes of 10 spherical harmonics, we need 10 calibration antennas whose spherical harmonic spectra are known, whose vectors of spherical harmonic amplitudes are linearly independent, and whose spectra illuminate the 10 spherical harmonics needed to properly reproduce an arbitrary half-wave dipole pattern. 
One possible choice for a set of calibration antennas is to use center-fed dipoles with various lengths and orientations. For simplicity, we consider ten identical half-wave dipoles but with different orientations, as shown in Figure \ref{fig:ref_dipoles}. 
The orientations of these calibration antennas were optimized to a local minimum of cond($\mathbf{A}$) using the MATLAB\textsuperscript{\textregistered} function fminsearch. 
The minimum value obtained was cond($\mathbf{A}$) = 41.22, 
which is still rather large considering that a well-conditioned matrix would have cond$(\mathbf{A}) \approx 1$.  
The larger condition number indicates that the matrix inverse process in Eq. (\ref{eq:QD_calc}) will be sensitive to small errors in $\mathbf{A}$, increasing the error of the pattern reconstruction. Clearly there is opportunity to improve the set of calibration antennas, but this set proves to be sufficient to illustrate the concept.

To calculate the spherical harmonic spectra of these antennas from first principles using Eqs. (\ref{eq:Mcoef}) and (\ref{eq:Ecoef}), we need the magnitude of the current at the antenna terminals, $I_0$, to determine the strength of the electric field. For simplicity, we will assume the antennas are fed with a lossless transmission line with characteristic impedance $Z_0$, and that the antennas are operated at resonance so that the terminal impedance $R_a$ is real. Under these assumptions the current at the terminals of the antennas is given by
\begin{equation}
    I_0 = \sqrt{\frac{2P_{inc}}{Z_0}}(1-\Gamma) = \frac{2|V_{inc}|}{R_a+Z_0},
\end{equation}
where $P_{inc}$ is the power incident from the source, $V_{inc}$ is the voltage of the incident wave on the transmission line to the antenna, and the reflection coefficient (assumed real as indicated above) is given by
\begin{equation}
    \Gamma = \frac{R_a-Z_0}{R_a+Z_0}.
\end{equation}
Note that if all of the calibration antennas and the test antenna have the same impedance, then the currents will all be the same and this calculation is not necessary. However, if any of the calibration/reference or test antennas have different terminal impedances, then the received signals need to be normalized to equivalent terminal currents in all of the antennas.

\subsection{Random Multipath Environment}
Let us model the voltage at the $k$th antenna probe as
\begin{dmath}
    v_k = \sum_{n=1}^{N_p} \rho_{k,n}\left[E_\theta (\theta_0,\phi_0,
    \theta_{k,n},\phi_{k,n})\cos\alpha_{k,n}\\ + E_\phi (\theta_0,\phi_0,
    \theta_{k,n},\phi_{k,n})\sin\alpha_{k,n}\right].
    \label{eq:v_rand}
\end{dmath}
Here $\rho_{n,k}$ represents the overall complex amplitude of the $n$th path to the $k$th antenna probe; $\theta_0,\phi_0$ represent the orientation of the axis of the dipole antenna being tested; $\theta_{k,n},\phi_{k,n}$ represent the launch direction from the test antenna of the $n$th path to the $k$th antenna probe; and $\alpha_{k,n}$ represents a polarization mixing angle owing to geometry and scattering along the $n$th path to the $k$th antenna probe. The parameters $\rho_{k,n},\theta_{k,n},\phi_{k,n},\alpha_{k,n}$ are taken to be random variables with distributions as described in Table \ref{tab:randparms}.
\begin{table}[]
    \caption{Parameters for the random multipath model}
    \label{tab:randparms}
    \centering
    \begin{tabular}{|c|l|}
    \hline
        Parameter & Distribution \\
        \hline
        $\rho_{k,n}$ & Normal (both real and imaginary parts),\\ 
        & $(\mu=0,\sigma_\rho$), $\sigma_\rho$ chosen to represent\\
        &received signal amplitude\\
        \hline
        $\theta_{k,n}$ & Uniform, $(0,\pi)$ \\
        \hline
        $\phi_{k,n}$ & Uniform, $(0,2\pi)$ \\
        \hline
        $\alpha_{k,n}$ & Uniform, $(0,2\pi)$ \\
        \hline
    \end{tabular}
\end{table}

This model was implemented with 10x10 random matrices for each of the path parameters in Table \ref{tab:randparms}, and $\sigma_\rho=0.001.$
The probe voltages $v_k$ were then calculated for each one of the reference antennas to create the square matrix $\mathbf{V}$. Since there is no physical model underlying the path parameter matrices, 100 sets of random path parameters were generated, and the set that yielded the lowest $\text{cond}(\mathbf{V})$ was kept for the demonstration model. The matrix $\mathbf{Q}\cdot\mathbf{D}$ was then calculated from Eq. (\ref{eq:QD_calc}). For the path parameters selected, $\text{cond}(\mathbf{V})=164.3$, and $\text{cond}(\mathbf{Q}\cdot\mathbf{D})=9.23$. 
Although the matrices were generated numerically from Eq. (\ref{eq:v_rand}) for the present example, it is also possible to obtain an analytical approximation for $\mathbf{Q}\cdot\mathbf{D}$ in terms of the model parameters, as shown in Appendix \ref{app:C}.

Owing to errors from ill-conditioned matrices (or experimental error in the case of measurements), the property (\ref{eq:coef_sym}) may not be satisfied by the direct calculation from Eq. (\ref{eq:a_calc}). Consequently, this property was imposed numerically using the calculation
\begin{align}
    \hat{a}_{l,m} &= \frac{1}{2}\left[a_{l,m}+(-1)^m a_{l,-m}^*\right],\\
    \hat{a}_{l,-m} &= (-1)^m \hat{a}_{l,m}^*,
\end{align}
where the directly-calculated values are $a_{l,m}$ and corrected values are denoted by $\hat{a}_{l,m}$.

\begin{figure}
    \centering
    \includegraphics[width=1\linewidth]{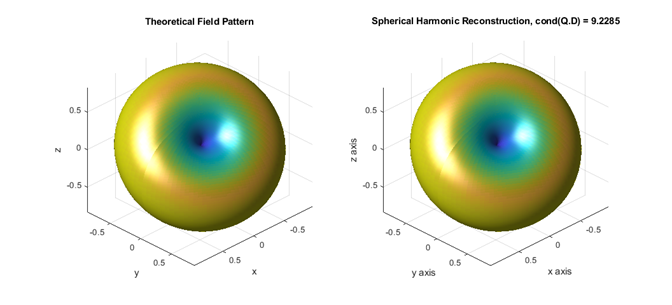}
    \caption{Example of simulated reconstruction in a random multipath environment.}
    \label{fig:EgReconstruct}
    \end{figure}

An example reconstruction for a test antenna orientation of $\theta_0 = \pi/4,\,\phi_0 = \pi/3$ is shown in Figure \ref{fig:EgReconstruct}, and the resulting estimates of field magnitude error, radiation resistance, and directivity are shown in Table \ref{tab:EgError}. The RMS field error was calculated by taking the difference between the theoretical and reconstructed field magnitudes at each sampled direction, normalizing the difference by the maximum magnitude of the theoretical field magnitude, and then calculating the square root of the mean of this normalized difference over all the sampled directions. 

    \begin{table}[]
        \caption{Example test case compared with theory 
        }
        \label{tab:EgError}
        \centering
        \begin{tabular}{|c|c|c|c|}
        \hline
             & RMS Field Error& $R_r \quad(\Omega)$& Directivity \\
             \hline
           Theory  & - & 73.1 & 1.64 (2.15 dB)\\
           \hline
           Reconstruction & & & \\
           $\theta_0 = \pi/4,\, \phi_0 = \pi/3$ & $7.86\times 10^{-4}$ &73.1 & 1.64 (2.15 dB)\\
           \hline
        \end{tabular}
    \end{table}

    \begin{figure}
        \centering
        \includegraphics[width=\linewidth]{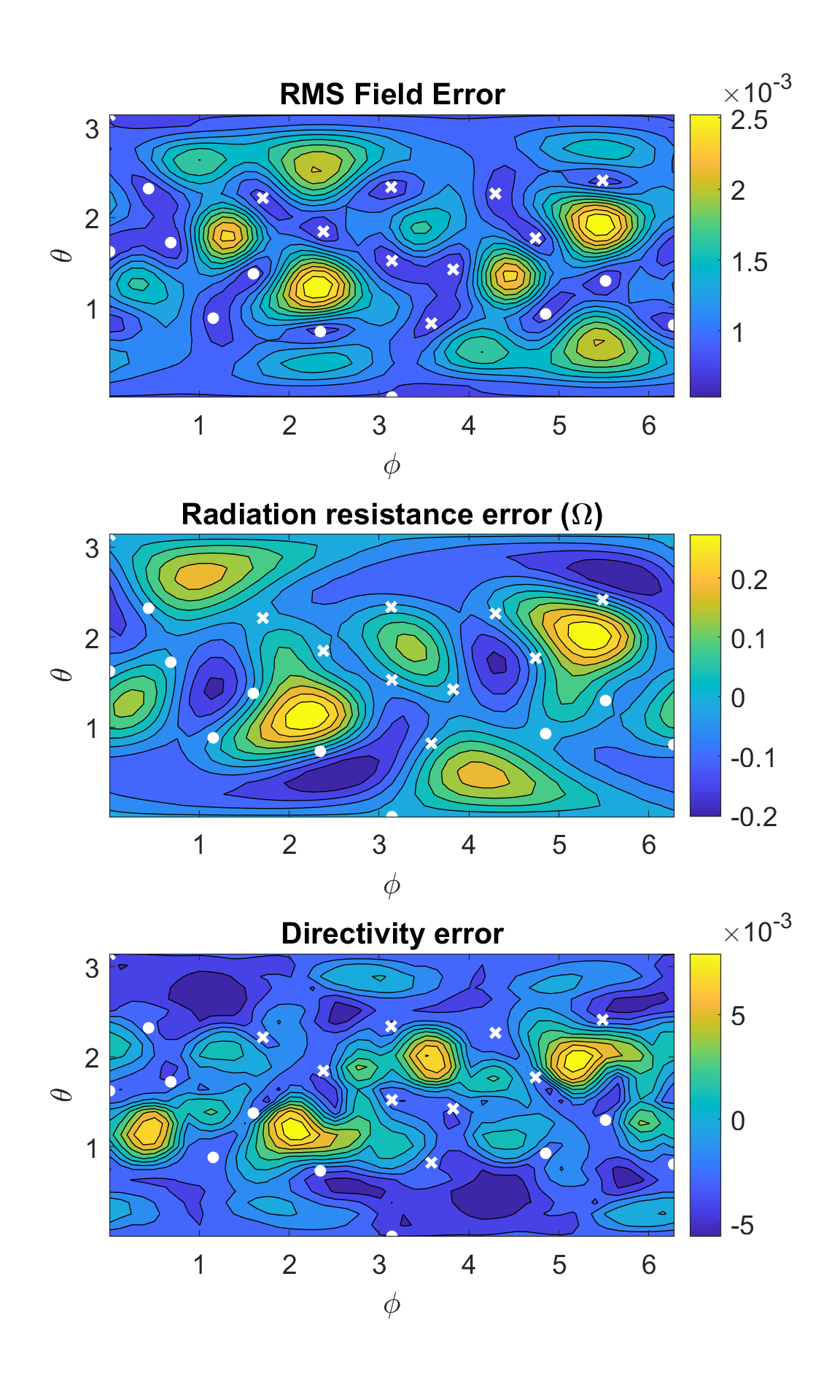}
        \caption{Summary of the errors in directivity and radiation resistance over the full range of test antenna orientations. The white circles indicate the positive axis of the calibration dipoles, and the crosses denote the negative axes (diametrically opposite to the positive axes). Note that the RMS field error is smallest near the reference antenna orientations, as expected. }
        \label{fig:EgFullError}
    \end{figure}
    The accuracy of the reconstruction can be seen to be quite good. This is to be expected since the only source of error is numerical error from the MATLAB\textsuperscript{\textregistered} computation. The errors in field magnitude, radiation resistance, and directivity over the full range of test antenna orientations are shown in Figure \ref{fig:EgFullError}.
    The radiation resistance and directivity errors are simply the difference between the theoretical values and the values obtained from the reconstructed pattern.

    Note that each set of random parameters can be viewed as corresponding to a particular set of probe antenna locations in the environment. It follows that some configurations of antennas give better-conditioned matrices than others. In an experimental facility designed to use this method for constructing patters, attention must be given to optimizing the placement of the probe antennas.

\section{Summary and Conclusions}
This work introduced a technique for measuring antenna patterns that constructively uses multipath propagation in the environment. As a result, the need for costly anechoic chambers is eliminated. Vector Spherical Harmonics are used as an efficient representation of the pattern, minimizing the number of required measurements and reference antennas. Example calculations to illustrate the method have been presented. An experimental demonstration of the technique is presented in a companion paper.

\begin{appendices}
\section{Expressions for Vector Spherical Harmonics}
\label{app:A}
\begin{align}
    \mathbf{X}_{l,m} &= \frac{-j}{\sqrt{l(l+1)}}\left[\hat{\phi}\sqrt{\frac{(2l+1)(l-m)!}{4\pi(l+m)!}}\right.\nonumber\\
    &\quad\times\left[\frac{l\cos{\theta}P_l^m(\cos\theta)-(l+m)P_{l-1}^m (\cos\theta)}{\sqrt{1-\cos^2\theta}}\right]e^{jm\phi}\nonumber\\
    &\quad -\left.\hat{\theta}\frac{jm}{\sin\theta}Y_{l,m}\right],
\end{align}
\begin{align}
    \hat{\mathbf{r}}\times\mathbf{X}_{l,m}&= \frac{j}{\sqrt{l(l+1)}}\left[\hat{\theta}\sqrt{\frac{(2l+1)(l-m)!}{4\pi(l+m)!}}\right.\nonumber\\
&\quad\times\left[\frac{l\cos{\theta}P_l^m(\cos\theta)-(l+m)P_{l-1}^m (\cos\theta)}{\sqrt{1-\cos^2\theta}}\right]e^{jm\phi}\nonumber\\
    &\quad +\left.\hat{\phi}\frac{jm}{\sin\theta}Y_{l,m}\right].
\end{align}
\section{Electric Field of Arbitrarily-Oriented Dipole}
\label{app:B}
Referring to the geometry shown in Figure \ref{fig:dipole}, the electric field from a center-fed dipole of arbitrary length and orientation is given by
\begin{align}
    \mathbf{E} &= -j\frac{\eta_0 I_0 e^{-jkr}}{2\pi r}(\hat{\boldsymbol{\theta}}p + \hat{\boldsymbol{\phi}}q)\nonumber\\
    &\quad\times\left[\frac{\cos{(kLg/2)}-\cos{(kL/2)}}{1-g^2}\right],
    \label{eq:E_arb}
\end{align}
where
\begin{align}           p&=\sin{\theta_0}\cos{\phi_0}\cos{\theta}\cos{\phi} +\sin{\theta_0}\sin{\phi_0}\cos{\theta}\sin{\phi\nonumber}\\
&\quad - \cos{\theta_0}\sin{\theta},\\
    q&=\sin{\theta_0}\sin{\phi_0}\cos{\phi}-\sin{\theta_0}\cos{\phi_0}\sin{\phi}\\
    g &=\sin{\theta_0}\cos{\phi_0}\sin{\theta}\cos{\phi}+\sin{\theta_0}\sin{\phi_0}\sin{\theta}\sin{\phi},\nonumber\\
    &\quad+\cos{\theta_0}\cos{\theta}.
\end{align}


\section{$\mathbf{Q}\cdot\mathbf{D}$ from Random Path Model}
\label{app:C}
Here we show that  it is possible to analytically construct an approximation to $\mathbf{Q}\cdot\mathbf{D}$ for the random path model.
Since there are only electric dipole components of the antennas, Eq. (\ref{eq:E_expand}) simplifies to
\begin{align}
    \mathbf{E}(\theta_0,\phi
_0,\theta,\phi) &=-\eta_0\sum_{l,m}(j)^{l+1}\,a_{l,m}^E\,\mathbf{\hat{r}}\times\mathbf{X}_{l,m}\nonumber\\
&=-\eta_0\sum_{l,m}(j)^{l+1}\,a_{l,m}^E\left[\hat{\theta}\left(\mathbf{\hat{r}}\times\mathbf{X}_{l,m}\right)_\theta\right.\nonumber\\
&\quad+\left.\hat{\phi}\left(\mathbf{\hat{r}}\times\mathbf{X}_{l,m}\right)_\phi\right]
\end{align}
or
\begin{align}
\mathbf{E(\Omega_0,\Omega)}&\equiv-\eta_0\sum_{q=1}^{N_A}(j)^{l_q+1}a_q^E(\Omega_0)\left[\hat{\theta}\,\mathbf{\Theta}_{l_q,m_q}(\Omega)\right.\nonumber\\
&\quad+\left.\hat{\phi}\,\mathbf{\Phi}_{l_q,m_q}(\Omega)\right].
\label{eq:E_rand}
\end{align}
Here we have replaced the two-dimensional sum with a one-dimensional sum using the mapping shown in Table \ref{tab:q_map}; explicitly indicated that the dependence of the dipole orientation $\Omega_0$ is contained in the spherical harmonic amplitudes while the dependence on the path launch angle $\Omega$ is contained in the spherical harmonic functions; omitted the pre-factor in view of the complex amplitude $\rho_{m,n}$; and we have introduced the short-hand notations
\begin{align}
    \theta,\phi\, &\rightarrow\,\Omega\\
    \left(\mathbf{\hat{r}}\times\mathbf{X}_{l_q,m_q}\right)_\theta&\rightarrow\Theta_{l_q,m_q}(\Omega)\\ \left(\mathbf{\hat{r}}\times\mathbf{X}_{l_q,m_q}\right)_\phi&\rightarrow\Phi_{l_q,m_q}(\Omega)
\end{align}

\begin{table}[]
    \caption{Mapping from the sum over $(l,m)$ to a single sum over $q$}
    \label{tab:q_map}
    \centering
    \begin{tabular}{|c|c|c|}
    \hline
        $q$ & $l$ & $m$ \\
        \hline
        1 & 1 & -1 \\
        \hline
        2 & 1 & 0 \\
        \hline
        3 & 1 & 1 \\
        \hline
        4 & 3 & -3 \\
        \hline
        5 & 3 & -2 \\
        \hline
        6 & 3 & -1 \\
        \hline
        7 & 3 & 0 \\
        \hline
        8 & 3 & 1 \\
        \hline
        9 & 3 & 2 \\
        \hline
        10 & 3 & 3 \\
        \hline
    \end{tabular}
\end{table}
Further, we have omitted the factor $\exp{(-jkr)}/(kr)$ in both (\ref{eq:E_expand}) and (\ref{eq:AE_coeff}), since this factor cancels when the equations are combined.

Substituting (\ref{eq:E_rand}) into (\ref{eq:v_rand}) allows us to write
\begin{align}
    v_k &= -\eta_0\sum_{q=1}^{N_A}a_q^E\left(\Omega_0\right)(j)^{l_q+1}\nonumber\\  &\quad\times\sum_{n=1}^{N_p}\rho_{k,n}\left[\Theta_{l_q,m_q}\left(\Omega_{k,n}\right)\cos{\alpha_{k,n}}\right.\nonumber\\
    &\quad +\left. \Phi_{l_q,m_q}\left(\Omega_{k,n}\right)\sin{\alpha_{k,n}}\right]\nonumber\\
&\equiv\left[\mathbf{Q}\cdot\mathbf{D}\cdot\mathbf{a}\right]_k,
\label{eq:v_approx}
\end{align}
where
\begin{align}
    \left[\mathbf{Q}\cdot \mathbf{D}\right]_{k,q} &=\nonumber\\
    &\quad - \eta_0 (j)^{l_q+1}\sum_{n=1}^{N_p}\rho_{k,n}\left[\Theta_{l_q,m_q}\left(\Omega_{k,n}\right)\cos{\alpha_{k,n}}\right.\nonumber\\
    &\quad +\left. \Phi_{l_q,m_q}\left(\Omega_{k,n}\right)\sin{\alpha_{k,n}}\right].
\end{align}
Note that estimating $v_k$ using Eq.(\ref{eq:v_approx}) will not yield exactly the same result as Eq. (\ref{eq:v_rand}) if the exact expression for the electric field is used, since (\ref{eq:v_approx}) uses an approximation to the electric field obtained by truncating the vector spherical harmonic series.

\end{appendices}



\bibliographystyle{IEEEtran}
\bibliography{IEEEabrv,myzotero3}

\begin{IEEEbiography}[{\includegraphics[width=1in,height=1.25in,clip,keepaspectratio]{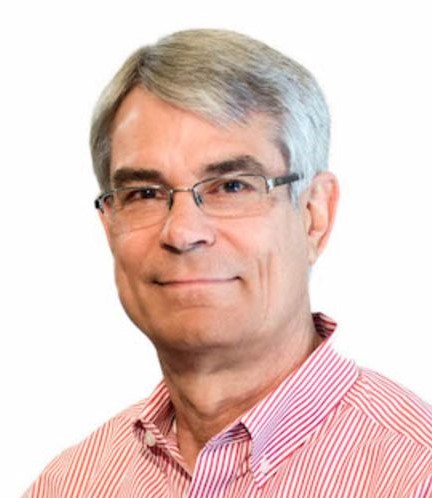}}]{Daniel D. Stancil} (S'75--M'81--SM'91--F'04--LF'20) received the B.S. degree in electrical engineering from Tennessee Technological University, Cookeville, in 1976 and the S.M., E.E., and Ph.D. degrees in electrical engineering from the Massachusetts Institute of Technology (MIT), Cambridge, in 1978, 1979, and 1981, respectively.  

From 1981 to 1986, he was an Assistant Professor of Electrical and Computer Engineering at North Carolina State University. From 1986 to 2009, he was an Associate Professor, then Professor of Electrical and Computer Engineering at Carnegie Mellon University, Pittsburgh, PA. At CMU he served as Associate Department Head of the ECE Department from '92-'94, and Associate Dean for Academic Affairs in the College of Engineering from '96-'00. He returned to North Carolina State and served as Head of the Electrical and Computer Engineering Department from 2009-2023. From 2019-2024 he was the Executive Director of the IBM Q Hub at NC State, a collaboration with IBM on quantum computing. He became Alcoa Distinguished Professor Emeritus in 2024. His research has included such varied topics as spin waves, optics, microwaves, wireless channels, antennas, remote labs, and particle physics.  

Technology for distributing wireless signals through HVAC ducts that Dr. Stancil and his students developed has been installed in such major buildings as Chicago's Trump Towers and McCormick Place Convention Center. The demonstration of neutrino communications by a multidisciplinary team coordinated by Dr. Stancil was recognized by Physics World Magazine as one of the top 10 Physics Breakthroughs of 2012. Additional recognitions that his work has received have included an IR 100 Award and a Photonics Circle of Excellence Award. Dr. Stancil has served as president of the IEEE Magnetics Society and president of the ECE Department Heads Association.
\end{IEEEbiography}


\end{document}